\documentclass[conference]{IEEEtran}
\usepackage{cite}
\usepackage{amsmath,amssymb,amsfonts}
\usepackage{algorithmic}
\usepackage{graphicx}
\usepackage{textcomp}
\usepackage{xcolor}
\usepackage{booktabs}
\usepackage[hyphens]{url}
\usepackage{hyperref}

\def\BibTeX{{\rm B\kern-.05em{\sc i\kern-.025em b}\kern-.08em
    T\kern-.1667em\lower.7ex\hbox{E}\kern-.125emX}}

\title{A Few Fit Most: Improving Performance Portability of SGEMM on GPUs using Multi-Versioning}

\author{
    \IEEEauthorblockN{Robert Hochgraf}
    \IEEEauthorblockA{
        \textit{Rochester Institute of Technology} \\
        Rochester, NY, USA
    }
    \and
    \IEEEauthorblockN{Sreepathi Pai}
    \IEEEauthorblockA{
        \textit{University of Rochester} \\
        Rochester, NY, USA
    }
}

\begin{document}
\maketitle
\thispagestyle{plain}
\pagestyle{plain}
\newcommand{\Description}[1]{{}}


\begin{abstract}
Hand-optimizing linear algebra kernels for different GPU devices and applications is complex and labor-intensive. Instead, many developers use automatic performance tuning (autotuning) to achieve high performance on a variety of devices. However, autotuning ``overfits", and must be redone if any part of the environment changes, such as if the device or input characteristics change.

In most non-trivial cases, a single compute kernel cannot maintain near-optimal performance across all environments. 
Changing the kernel to specialize it to the current execution environment is possible, but on GPUs, runtime tuning and compilation can be expensive.

In this work, we use \textit{multi-versioning} -- producing several variants of the same code -- as a way to generate performance portable code. 
We describe a framework called \textit{portability tuning} that can automatically generate multi-versioned code whose performance is portable, requiring no retuning.

We evaluate our framework on a dataset of execution times for GEMM kernels from the CLBlast linear algebra library. 
We find our portability tuning techniques outperform CLBlast's default kernels -- often approaching within 10\% of the theoretical maximum performance -- despite CLBlast using autotuning techniques. 
Further, we find that our generated programs generalize well to new and unseen devices, matching the performance of autotuning \textit{without ever portability tuning for those devices}.
\end{abstract}

\section{Introduction}

The issue of \textit{performance portability} -- achieving high performance among diverse execution environments -- is a pressing concern for developers of high-performance computing (HPC) applications, libraries, and frameworks. 
As architectural diversity increases, developing code that maintains performance efficiency among various platforms is increasingly challenging~\cite{Balaprakash, Deakin2019, Kokkos}.

In order to develop libraries and applications whose performance is portable, some developers have turned to autotuning. In this technique, performance-critical portions of code called \textit{kernels} are parameterized with the parameters controlling code features such as tiling sizes and loop unrolling factors. Autotuning then determines the optimal parameters for the given architecture and input, usually through empirical search. 

Unfortunately, autotuning often \textit{overspecializes} kernels to the original execution environment. If run for another device or input, the tuned code may inefficiently utilize the hardware~\cite{Price}.
For example, on the Mali-G71, the tuned default matrix-multiply kernel for CLBlast~\cite{10.1145/3204919.3204924} achieves less than 15\% of optimal performance for certain sizes of input matrices.\footnote{In our experiments, for (M,N,K) = 512x4096x4096, 4096x4096x512, 1024x1024x4096, 4096x1024x4096.} This is because the default matrix-multiply kernel was fully specialized for multiplying two 1024x1024 matrices. Similar slowdowns can be noted when porting the kernel to other devices.

To achieve performance portability through autotuning, it needs to be repeated for each intended execution environment (GPU device and program input) so that fast parameters are known for each possible invocation. Then, for example, a model-based system could be used to automatically select and run the tuned code at runtime~\cite{Muralidharan}.

However, the model-based solution is hard to implement when the number of parameter configurations is large, as is the case for complex kernels such as matrix multiplication. 
As we will show in Section~\ref{Observations and Motivation}, runtime compilation is expensive.
To avoid runtime compilation, a code variant must be compiled ahead of time, adding to the library size and posing a significant storage cost.

A restricted set of variants could be compiled ahead of time, but these would now need to be identified.
Lacking a clear methodology to identify performance-portable code variants, state-of-the-art methods often rely on hand-selection. 
NVIDIA's cuBLAS library contains 24 pre-selected GEMM kernels, for example~\cite{cublas_2022}. 
However, hand-selection requires intensive developer effort and does not scale as the number of devices and inputs increase. 

In this work, we address the problem of automatically generating performance portable programs by identifying semi-specialized code variants that provide high performance on a space of execution environments. We demonstrate the practical tradeoffs between performance portability and the number of generated variants, and provide a framework to automatically navigate these tradeoffs. Our framework can accommodate a number of different kernel pre-selection mechanisms and performance goals.

We extensively evaluate our framework experimentally, by applying it to the matrix multiplication kernels, i.e. GEMM.
GEMM kernels are often parameterized. Furthermore, the performance attained by GEMM kernel variants varies by environment in often confounding ways, experiencing various bottlenecks on the hardware. This complexity provides a rigorous test for our methodology.
Our evaluation space is broad, involving a kernel compiled for up to 1,775 unique parameter configurations, executed with up to 64 unique program inputs on 5 GPUs from 4 vendors. 
We consider and evaluate our framework for a variety of different implementations and highlight the tradeoffs posed between these implementation decisions. 

Our contributions can be summarized as:

\begin{itemize}
\item We present the portability tuning framework for automatically generating performance portable code. Portability tuning unifies prior work in this area, particularly several recent research directions in \textit{semi-specialization}, \textit{multi-versioning}, and \textit{input sensitivity}.
\item We build a large dataset on the performance of many GEMM variants on a diverse set of GPUs, executed with many input sizes that can be used for future work in portability tuning.
\item We evaluate several different implementations of this framework. For example, we show that \textit{any general-purpose autotuner} is able to generate performance portable code using our framework.
\item We demonstrate that the generated code variants remain performance portable even onto \textit{unseen} GPUs and inputs. 
\end{itemize}

The paper is organized as follows: in Section~\ref{sec:related}, we describe related work. In Section \ref{Observations and Motivation}, we motivate the problem of \textit{kernel pre-selection} which informs our design of the portability tuning framework proposed in Section \ref{Portability Tuning Framework}. In Section~\ref{Experimental Methods / Dataset Generation}, we generate a dataset on which we evaluate our proposed framework and evaluate portability tuning for a variety of different tuning cases. We 
conclude in Section~\ref{sec:conclusion}.

\section{Related Work}
\label{sec:related}

\begin{table*}
\small
  \caption{Related Work Comparison}
  \label{tab:related work comparison}
\centering
\begin{tabular}{|c|c|c|c|c|c|c|c|}
\hline
  \textbf{Authors} & \textbf{Method} & \textbf{Input Portable?} & \textbf{Device Portable?} & \textbf{Multi-Version?} & \textbf{Performance Portable?} \\
\hline
\textbf{Sorensen et al.} & Rank Algorithm  & $\checkmark$  & $\checkmark$  & $\times$  &  $\checkmark$  \\
\hline
\textbf{Price and McIntosh-Smith} & Autotuning  &  $\times$ & $\checkmark$ & $\times$  & $\checkmark$  \\
\hline
\textbf{Tillet and Cox} & MLP  & $\checkmark$  &  $\times$ &  $\times$ & $\times$  \\
\hline
\textbf{Labini et al.} & Decision Tree  & $\checkmark$  & $\times$  & $\times$ &  $\times$ \\
\hline
\textbf{Lawson and Goli} & Clustering  &  $\checkmark$ &  $\times$ & $\checkmark$ & $\checkmark$  \\
\hline
\textbf{This work} & \textit{PortabilityTune}  &  $\checkmark$ &  $\checkmark$ & $\checkmark$ & $\checkmark$  \\
\hline
\end{tabular}
\end{table*}

Prior work has investigated autotuning for a variety of applications, including Stencil calculations~\cite{5470421}, sparse matrix-vector operations~\cite{RichardVuduc2005}, dense linear algebra~\cite{CLINTWHALEY20013}, and parallel data compression and I/O~\cite{9671876}. A number of general-purpose tuning frameworks have been proposed to perform these tunings, including OpenTuner~\cite{OpenTuner}, clTune~\cite{clTune}, GEIST~\cite{GEIST}, GPTune~\cite{GPTune}, Kernel Tuner~\cite{kerneltuner}, BLISS~\cite{Bliss}, and ATF~\cite{ATF}. 
These frameworks use a variety of search methods to determine high-performing code configurations while only requiring the ability to time a program.

Autotuning been proposed as a way of achieving performance portability. However, tuned code exhibits low performance portability~\cite{Price, Sorensen, Labini2021}, so this proposition intends that autotuning be re-run for each intended execution environment, which is often expensive. Additionally, tuned code may not retain performance on other environments, so even minor changes applications may lead to significant performance degradation~\cite{Balaprakash}. Some works have proposed "transfer tuning" \cite{GPTune} approaches that model high-performing areas of the parameter space from prior performance results, significantly decreasing the number of evaluations necessary during tuning. Others use predictive models to directly predict parameters \cite{Labini2021, Tillet}. Though mitigating the cost of the tuning itself, re-tuning can introduce significant additional costs, as we discuss in Section \ref{Observations and Motivation}.

Prior work has introduced several strategies for addressing the heterogeneity introduced by varying hardware, applications, and program inputs. 
In general, this work aligns along four directions: 

\begin{enumerate}
    \item Determining code transformations whose performance generalizes (``performance portable``).
    \item Adapting code for a given architecture (``device sensitivity").
    \item Adapting code for a given input (``input sensitivity").    
    \item Selecting a set of code variants (``multi-versioning").
\end{enumerate}

Of these four directions for achieving performance portability, prior work typically considers only two as summarized in Table~\ref{tab:related work comparison}, and described in more detail below.

Sorensen et al.~\cite{Sorensen} propose an algorithm to \textit{semi-specialize} compiler optimizations for graph algorithms.
Using a rank-based statistical test, the algorithm yields a single set of compiler flags specialized for chosen set of dimensions from architecture, input, and application. 

\textit{PortabilityTune} can similarly generate semi-specialized parameters subsuming this work, and we explore this for more complex environmental factors, not just a set of compiler flags.
Additionally, our work expands semi-specialization by considering semi-specialization with multi-versioning. 
Through piece-wise combination of several semi-specialized kernels, we show that the achievable performance portability of programs increases.

Price and McIntosh-Smith~\cite{Price} evaluate the performance portability of custom Jacobi solver kernels implemented in OpenCL by autotuning the kernels on ten devices and assessing their performance when executed on the other devices. 
It concludes that the tuned kernels are over-specialized for each device and thus may lose significant performance if run on another device. To remedy this, the work proposes a multi-objective autotuner that tunes across multiple devices simultaneously and tunes a single kernel for all ten devices. 
The resulting single kernel delivers much higher performance portability. 
Our work considers the impact of input heterogeneity and multi-versioning, and can use different fitness functions to meet various performance goals.

Several works investigate direct prediction of kernel parameters with machine learning, such as for sparse matrix-vector operations~\cite{osti1367962, 7965113}.
Tillet and Cox~\cite{Tillet} propose a framework for input-aware autotuning called ISAAC.
It uses autotuning to collect performance data for many program inputs. 
Then a predictive model is trained from the execution data to directly infer kernel parameters for a given input.
ISAAC uses a multi-layer-perceptron model to predict parameters for parameterized PTX kernels. Labini et al.~\cite{Labini2021} similarly construct and evaluate the performance of several machine learning models for addressing input sensitivity in GEMM kernels.
In contrast to these methods, which generate highly specialized code on a per-architecture and per-input basis, our work considers the generation of kernels whose performance is portable across environments. By doing this, we mitigate drawbacks from re-compilation and storage, which we discuss in Section \ref{Observations and Motivation}.

A series of prior work has investigated clustering strategies to determine the set of high-performance kernel variants and provided methodologies for handling input sensitivity, such as a two-level framework for algorithmic choice~\cite{10.1145/2737924.2737969} and tree-based strategies for compilers of parallel codes~\cite{10.1007/978-3-030-83978-9_1, 10.1145/3293883.3295707}. 
Of these approaches, we are closest to Lawson and Goli~ \cite{LAWSON2021102813} which addresses input sensitivity and multi-versioning for a GEMM kernel implemented in SYCL. 
That work uses clustering methods to prune the space of possible kernels into a small set that preserves a large proportion of the potential performance. Then a  decision tree selects between these kernels at runtime based on the incoming input. 
We evaluate clustering methods as part of our framework for the case of device sensitivity on complex, non-standardized datasets and find that the method may not reach the highest-performing selections.

Prior work also investigates the use of multi-versioning to improve the performance of input-sensitive code. In multi-versioning, each code version is guarded by predicates so that the optimal version is selected at runtime on each run~\cite{10.1145/3293883.3295707}. This may provide a fast and storage-efficient way to provide several code variants and achieve high performance portability. Prior works have proposed several frameworks to automatically multi-version input-sensitive code, however these methods focus on optimizing parallelism and may not be suitable for determining tile sizes, whose selection is key to achieving high performance with linear algebra routines such as matrix multiplication~\cite{10.1007/978-3-030-83978-9_1, 10.1145/3204919.3204924}.

Our work seeks to unify these four directions. 
We propose a framework that automatically selects a set of semi-specialized code variants for a given space of devices, inputs, and other environmental factors.
These selected kernels balance performance, portability, and the number of code variants.

\section{Observations and Motivation}
\label{Observations and Motivation}

We first define the terminology used in this paper. Then, we present drawbacks to autotuning GPU kernels in context of the tradeoffs posed by performance portability. Lastly, we present an intuition for how \textit{kernel pre-selection} and \textit{multi-versioning} can be used to increase the performance portability of programs.

\subsection{Terminology}

\noindent \textbf{Tuning parameters}: program-level preprocessor constants that affect program characteristics such as workgroup sizes and thread utilization. The values of the parameters affect the execution time of the program.

\noindent\textbf{Parameter configuration} (or \textit{parameter variant}): a single combination of the tuning parameters.

\noindent\textbf{Code variant} (or \textit{kernel variant}): a kernel compiled with some parameter combination.

\noindent\textbf{Execution environment} (or \textit{environment}): a tuple of the \textit{input} (e.g., matrices) provided as arguments to the kernel and a GPU \textit{device} that kernel executes on.

\noindent\textbf{Performance portability}: a measurement of the expected performance of a program on the execution environments of interest.

\noindent\textbf{Code divergence}: the number of code variants in a multi-versioned program.\footnote{In Pennycook et al.~\cite{PENNYCOOKNAVIGATINGPERFPORT}, \textit{code divergence} is a measure of developer effort, representing the amount of environment-specific code (since each environment-specific code requires separate maintenance). Here, we consider \textit{code divergence} as the number of code variants, imposing a cost in terms of storage (if precompiling kernels) or performance overhead (if recompiling at runtime).}

\subsection{Challenges to Achieving Performance Portability on GPUs}
\label{Challenges to Achieving Performance Portability on GPUs}

Pennycook et al.~\cite{PENNYCOOKNAVIGATINGPERFPORT} motivate the issue of performance portability as a tradeoff between performance portability and code divergence.

At one end of this tradeoff, there may exist a parameter configuration for which a kernel would achieve peak performance for any environment (GPU device and input) of interest. This corresponds to maximal performance portability and minimal code divergence. However, such parameters do not exist in most non-trivial code~\cite{Sorensen}. Thus, to achieve higher performance portability, the use of multi-versioning (via multiple parameter variants) is necessary.

At the other end of the tradeoff, a multi-version approach can guarantee high performance portability by autotuning to determine a high-performing parameter configuration whenever the execution environment changes. This  generates a high number of tuned codes. On GPUs, we find that this approach to autotuning is limited by the execution model. Since the execution environment may not be known until execution has begun, (1)~a (near-)optimal parameter configuration must be determined at runtime, and (2)~a kernel variant compiled with the aforementioned parameter configuration must be available for execution at runtime.

Model-based autotuning systems such as ISAAC~\cite{Tillet} can directly determine and predict parameters to avoid autotuning at runtime. But this requires that prediction models be available for each intended GPU device.
Having a kernel variant with the required parameter configuration available is more challenging.
The two approaches possible are \textit{precompiling} all possible kernel variants and \textit{recompiling} kernels with the selected parameter configuration at runtime prior to execution.
We show that neither is viable in the following sections.

\subsubsection{Costs of Recompilation}
\begin{table}
\centering
  \caption{GEMM: Runtime vs Compilation (ms)}
  \label{tab:gemm compilation}
\begin{tabular}{cccc}
\toprule
Device & Median Runtime & Median Compile Time & Slowdown \\
\midrule
Vega & 1.92 & 266.47 & 138.6x \\
Iris Pro & 19.45 & 201.42 & 10.4x \\
HD500 & 39.49 & 736.13 & 18.6x \\
Quadro & 0.89 & 792.44 & 894.4x \\
Mali & 95.00 & 683.23 & 7.2x \\
\bottomrule
\end{tabular}
\end{table}
Prior work in GEMM parameter inference does not consider the time for kernel recompilation \cite{Labini2021, Tillet}. We investigate the just-in-time compilation times for the OpenCL kernels provided in the CLBlast library.

To change between kernels at runtime, CLBlast provides the \textit{OverrideParameters} API call. This function overrides the kernel parameters so that on the next kernel call, the kernel will be re-compiled with new performance characteristics. We quantify the compilation costs for the multiplication kernel Xgemm in Table \ref{tab:gemm compilation}. Our results show that in most cases, the time to recompile a kernel is much greater than that of running that kernel. For example, on the NVIDIA Quadro device, the median compilation time of the Xgemm kernel was comparable to the runtime of nearly 900 kernel executions. Because of the compilation overhead, methods that rely on runtime compilation face significant costs that hamper overall performance.

\subsubsection{Costs of Precompilation}
\begin{table}
\centering
  \caption{GEMM Kernel Sizes (KB)}
  \label{tab:gemm kernel sizes}
\begin{tabular}{ccccccc}
\toprule
    Device & Vega & Quadro & HD500 & Iris Pro & Mali \\
\midrule
    Avg. Size & 14.74 & 7.32 & 22.36 & 21.62 & 10.26 \\
\bottomrule
\end{tabular}
\end{table}
Runtime compilation can be avoided by precompiling (caching) kernel binaries. Precompilation may even be necessary in frameworks like SYCL that deploy using compiled binaries. However, precompilation poses costs in terms of program size and build time. 
In Table~\ref{tab:gemm kernel sizes}, we report the average kernel sizes for the compiled GEMM kernels on each device from a sample of 100 GEMM kernels with a variety of parameters. 
While a single compiled binary is not expensive to store -- taking just 7.3KB of space for the NVIDIA Quadro -- there are over a million kernel configurations for some devices. Storing all of the pre-compiled kernels would take on the order of gigabytes of storage. It would also take over 9 days to compile on our hardware. 
An informed and limited selection is necessary.

\subsubsection{Kernel Pre-Selection}

\begin{figure*}
    \centering
    \includegraphics[width=0.9\textwidth]{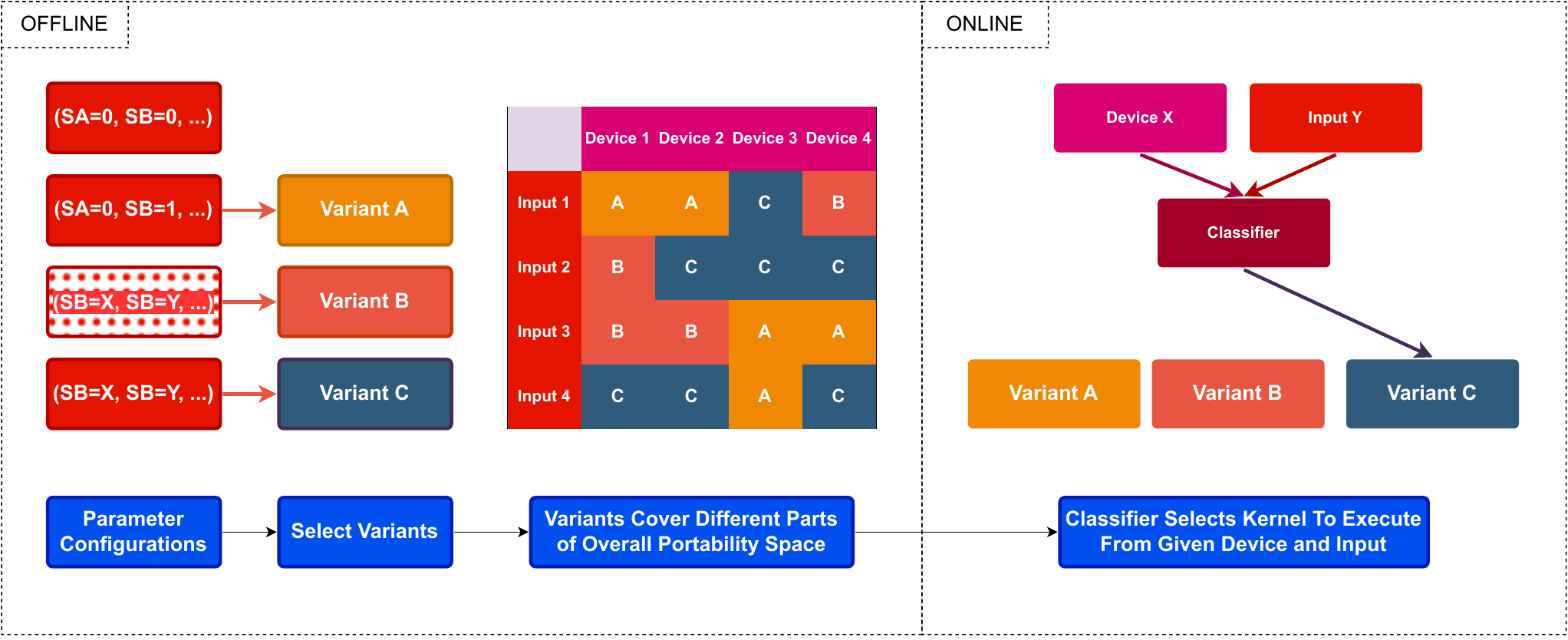}
    \caption{Kernel Pre-Selection and Runtime Inference}
    \Description{First, offline, kernels are selected from a large space of possible kernels. These kernels cover different parts of the overall space of devices and inputs. Then, online, a classifier determines which of the selected kernels should be executed from the given device and input.}
    \label{fig:The Portability Tuning Framework}
\end{figure*}

State-of-the-art vendor libraries often use an alternative approach. These libraries may perform \textit{kernel pre-selection} -- designing a small set of kernels, from which corresponding heuristics can select between at runtime.\footnote{For example, the state-of-the-art cuBLAS has 24 pre-selected matrix multiplication variants~\cite{cublas_2022}.}

The above scenario is depicted in Figure \ref{fig:The Portability Tuning Framework}. From a large space of possible kernel implementations, three specific variants A, B, and C are chosen. These three variants can then be chosen to cover different parts of the space by a runtime model. Since all three variants can be executed for each execution environment, the performance for each environment is that of the best-performing of the three variants on that environment. The goal in pre-selection, then, is to select \textit{semi-specialized} variants that maximize the maximal performance of the selected variants on each environment.

Kernel pre-selection represents a middle-ground of the tradeoff between performance portability and code divergence. 
The resulting program remains multi-versioned, improving performance portability over a single-version approach -- but limits the number of variants, to lower build time and storage size costs.

It is common for vendor libraries to perform kernel pre-selection manually, which requires intense developer effort. 
Worse, the process can leave some execution environments for which the performance is poorly optimized~\cite{Tillet, Labini2021, Lime}.

In this work, we provide a method, analogous to autotuning, that automatically performs kernel pre-selection to automatically generate program code whose performance is portable.

\section{The Portability Tuning Framework} \label{Portability Tuning Framework}

The portability tuning framework we describe in this section automatically performs kernel pre-selection. Portability tuning expands autotuning from a function operating on individual devices, inputs, and kernels, to one that operates on sets of those items. By diversifying the space considered during tuning, the resultant variants are able to generalize significantly further, and thus may be more performance portable than those selected by autotuning.

As a framework, we discuss several implementation options possible within portability tuning based on existing approaches, such as with general-purpose autotuners and machine learning techniques. Lastly, we discuss the performance evaluation of the selected kernel variants.

\subsection{Intuition}

A major limitation of autotuning is that autotuning overspecializes kernel parameters for a specific input, device, and environment. Formally, let $\mathit{AutoTune}$ to be a function that picks a kernel $k$ that is the best for input $i$, device $d$ and other environmental factors $e$ from among a set $K$ of kernel variants:
$$
k = \mathit{AutoTune}(i, d, e)
$$

Here, $i \in I$ where $I$ is the set of all possible inputs, $d \in D$ with $D$ the set of all devices, and $e \in E$ where $E$ is set of environments that are intended to model other aspects of the system that influence performance such as operating system, compiler, and so on.  

\textit{AutoTune} is limited by the intuition that we can only guarantee performance of a given kernel for a specific platform.
However, the kernel pre-selection process we discussed in Section \ref{Observations and Motivation} shows this is not a necessary constraint -- variants can be adapted for performance across environments.

\subsection{Proposed Method}

Our generalization $\mathit{PortabilityTune}$ instead returns a \textit{set} $\kappa$ of variants that are adjudged to be the best according to some metric for the \textit{sets} of inputs $\iota$, devices $\delta$, and environment $\epsilon$. 
$$
\kappa = \mathit{PortabilityTune}(\iota, \delta, \epsilon)
$$

Here, $\iota \in \mathcal{P}(I)$ (or equivalently, $\iota \subseteq I$), and similarly $\delta \in \mathcal{P}(D)$ and $\epsilon \in \mathcal{P}(E)$, where $\mathcal{P}(\cdot)$ indicates the power set of its argument. In practice, the size of $\kappa$ will also be constrained, in order to limit storage costs.

By relaxing the constraint found in autotuning to identify a single kernel, tuning can be more flexible and can now select a small number of \textit{semi-specialized} kernels, with each \textit{semi-specialized} kernel mapping to a portion of the overall space where it achieves high performance.

As an illustrative example, consider the case when $|\kappa|=1$, i.e. $\mathit{PortabilityTune}$ is constrained to return a single kernel variant, just like $\mathit{AutoTune}$.
Traditional autotuning would produce a tuned variant for a device and input pair in a particular environment, whereas portability tuning produces a variant tuned for a particular \textit{set} of execution environments. 
Essentially, a kernel is evaluated in all the environments, and the individual evaluations combined, to arrive at the kernel's performance portability.
Thus, the variant contained in $\kappa$ would be more performance portable than the variant $k$ returned by $\mathit{AutoTune}$ by construction.

As the size of $\kappa$ increases, the individual selected variants within $\kappa$ may become more specialized -- leading to increased performance -- but the overall set of variants $\kappa$ will still remain general over the examined execution environments. This process preserves the portability of the performance while still enabling the specialization necessary for better performance.

\subsection{Implementations}

\subsubsection{Autotuning} 
Portability tuning can be implemented using just a standard general-purpose autotuner. There are two problems that must be solved. 

First, the autotuner must search through the space of variant combinations $\mathcal{P}(K)$, the powerset of kernel variants.
Then, for each combination, the tuner must determine the fitness of the kernel combination according to a summary metric. At the end of the tuning, the autotuner returns the variant combination with the highest ranking.

While some of these spaces can be exhaustively searched, others are large enough to be intractable. The number of evaluations of device--input--environment--parameter variant(s) tuples is bounded by $O(|\iota|\times|\delta|\times|\epsilon|\times|K|^{|\kappa|})$, meaning that the computational cost can be significant, even compared to kernel executions (if building the dataset on-line).\footnote{If executing kernels for a portability tuning online, the number of kernel executions is bounded by $O(|\iota|\times|\delta|\times|\epsilon|\times|K|*n)$, where $n$ is the number of samples, assuming that the result of running a kernel on a device and input is stored and retrieved from a database.}

We found experimentally that for tunings with $|\kappa|>2$, the search space is often unrealistic to exhaustively evaluate. For example, there are 1,343 potential Xgemm parameter variants for the Quadro GPU in our dataset. To exhaustively evaluate the best 3 Xgemm kernels for the Quadro on 10 inputs would require over 24 billion (variant1, variant2, variant3)-input evaluations. 

In cases such as the scenario described above, heuristic and stochastic search approaches provided in state-of-the-art autotuners may help speed the tuning convergence~\cite{clTune, GPTune, kerneltuner}. Alternatively, machine learning methods can be used to prune the search space. We discuss two machine learning methods adapted from prior work \cite{LAWSON2021102813, Labini2021} next.

\subsubsection{Unsupervised Clustering}
One way to prune the search space is to perform unsupervised clustering on the data. In this work, we will consider the $k$-means clustering algorithm. $K$-means is a simple iterative algorithm that attempts to find a set of $k=|\kappa|$ centroids with minimal distance to the points in their cluster.

For a given execution environment, the dataset contains performance results for $|K|$ parameter configurations. We can represent the performance results for each execution environment as a point in $N$-dimensional space, where $N$ is the number of variants in the dataset. 
In this high-dimensional space, execution environments with similar performance results will be located near each other. 

This structure allows clustering to group the execution environments by their performance results and return representative centroids. From the performance results vector of the centroid, we select the highest-performing variant. By clustering, the dimensionality of the search is reduced from $|K|^{|\kappa|}$ to $|K|\times|\kappa|$.

\subsubsection{Supervised Machine Learning}
Another method to reduce the search space is to use supervised machine learning models. In this work, we consider the case of a decision tree. By training a decision tree as a regressor, the decision tree will estimate the performance vector for a given execution environment. Then, by limiting the tree to contain $|\kappa|$ leaf nodes, each leaf will contain a representative performance vector for the execution environments that end at that node in the tree. For each leaf node, a representative semi-specialized variant is chosen by the variant that provides the highest performance across the execution environments represented by the vector. Similar to clustering, this process reduces the dimensionality of the search from $|K|^{|\kappa|}$ to $|K|\times|\kappa|$.

\subsection{Evaluating the Selected Variants} \label{sec: Evaluating the Selected Variants}
So far, we have omitted how to evaluate the performance of the selected variants.
In autotuning, the performance of the parameter configuration is the fitness criterion.
For portability tuning, we evaluate a set of variants over a set of execution environments, so we also need a \textit{cost function} that summarizes results among many execution environments into a single fitness value. We also need to choose an appropriate \textit{performance metric} to be measured and summarized.

In the following sections, we derive the cost functions and performance metrics from the performance goals of library design and fleet optimization.

\subsubsection{Library Performance} Library designers often seek to design their library so that high overall performance is achieved across many environments.
For this goal, the choice of absolute runtimes as a performance metric and the mean as the cost function would be inappropriate. The resulting set of kernels would be highly biased towards devices and inputs with high runtimes, and may not provide high performance for devices and inputs that have shorter runtimes overall since the optimization criteria of minimizing kernel time would benefit most by focusing on the (absolutely) slow devices.

We can mitigate this bias by instead maximizing the performance for each device and input relative to an oracle, or best kernel, for that device and input. The overall performance can then be well summarized by the geometric mean. 

With this cost function and performance metric, the overall cost of a set of kernels $K$ tuned for a set of devices $D$ and inputs $I$ can be found with the following:

\begin{equation}
     \mathit{geomean(\{     \frac{\mathit{max}_{k\in{K}} {\mathit{runtime}(k)}_{d,i}}
                     {\mathit{oracle}_{d,i}} \mid d,i \in D,I\})}
\end{equation}    

Note that our choice is not free of biases. 
Devices have varying levels of sensitivity to tuning. For example, for the Quadro device, the middle 50\% of the variants have slowdowns of 1.8x to 4.3x, whereas for the Mali this is 2.6x to 13.2x. 

A device with high sensitivity -- such as the Mali -- will respond more to changes in the set of kernels. 
In practice, this means that the tuning will effectively weight sensitive devices and inputs over other devices and inputs, and thus favor those environments.

\subsubsection{Fleet Performance}

Distributed computing projects often coordinate a fleet of heterogeneous devices in order to finish tasks, where the goal of the fleet is to collectively complete as many tasks as possible. Portability tuning could be configured to select a single version of a library containing a handful of kernels, that maximizes the overall rate that the fleet completes tasks.
In this variant of portability tuning, unlike tuning for library performance, we seek to take advantage of the population characteristics of devices and inputs. 

A distributed computing fleet has varying numbers of each type of device. The ``tasks" of that fleet, while standard, also vary in how many of each input are used per task. We represent the quantity of devices $d$ in the fleet as $\textit{quantity}(d)$, the quantity of each input $i$ in the tasks as $\textit{quantity}(i)$, and the runtime of an input on a device as $y'$. Then, the rate the fleet completes tasks is equal to:
\begin{align}
\sum_{d \in D} \frac{\text{quantity}(d)}{\sum_{i \in I} y' * \text{quantity}(i)}
\end{align}

Recall that the cost function is minimized through the tuning.  Therefore, we can maximize the rate that tasks are completed by the fleet by minimizing its reciprocal. By setting the cost function and performance metric in this way, we enable portability tuning to directly optimize the rate at which tasks are completed by the fleet.

\section{Evaluation}
\label{Experimental Methods / Dataset Generation}
\label{Results/Discussion}

To evaluate portability tuning, we first generate a dataset of kernel execution times for CLBlast's matrix multiplication kernel (Xgemm) for different inputs (i.e. matrix sizes) on different GPUs~(Table~\ref{tab:gpus studied}).
The performance characteristics of kernels are determined by their \textit{parameters}, which define properties such as the tile sizes and loop unrolling factors. 
Apart from the data we collect (which contains 5 GPUs), we also reuse CLBlast's database on crowdsourced execution times, which after clean up, results in data from 71 GPUs.

\subsection{GPUs}

\begin{table}
\centering
  \caption{GPUs Studied}
  \label{tab:gpus studied}
  \begin{tabular}{ccccl}
    \toprule
    GPU & Vendor & \# CU & Market & Short Name  \\
    \midrule
    Mali-G71 & ARM & 12 &  Phone & Mali \\ \hline
    Iris Pro & Intel & 48 & NUC & Iris \\ \hline
    HD500 & Intel & 8 & NUC & HD500 \\ \hline
    RX Vega 64 & AMD & 64 & Desktop & Vega \\ \hline
    Quadro P5000 & NVIDIA & 20 & Workstation & Quadro\\
    \bottomrule
  \end{tabular}
\end{table}

The kernels were executed on 5 devices spanning 4 vendors: the Intel HD500, Intel Iris Pro, ARM Mali-G71, AMD Radeon RX Vega 64, and NVIDIA Quadro P5000. 
The NVIDIA and AMD devices are discrete GPUs, the others are integrated GPUs. 
The devices span several years (2015--2019) of make and have highly different performance capabilties, including both a mobile phone GPU (Mali) and a high-end workstation GPU (Quadro).
We investigate these GPUs since GPUs released from these years remain prominently used by consumers, making them important for our application domain of volunteer computing.\footnote{For example, over 1/3 of the top 100 most common GPUs on the gaming platform Steam in December 2023 were released between 2015 and 2019.~\cite{steamsurvey}}

\subsection{Inputs}

\begin{table}
\centering
  \caption{Inputs Evaluated}
  \label{tab:dataset generation}
  \begin{tabular}{cccl}
    \toprule
    Device & Number of Inputs & Number of Variants & Number of Entries   \\
    \midrule
    Mali & 63 & 1224 & 50472\\
    Vega & 64 & 1232 & 55043\\
    Quadro & 64 & 1343 & 62843\\
    Iris & 56 & 1207 & 55248\\
    HD500 & 44 & 1207 & 38490 \\
    \bottomrule
  \end{tabular}
\end{table}

GEMM operations multiply M*N and N*K matrices, resulting in a M*K matrix as an output. We execute the GEMM kernels for a regular set of input matrices where each M, N, K is one of $\{256,512,1024,4096\}$. This gives 64 total square and rectangular matrix multiplications. These varying settings test different performance characteristics of each kernel on the GPUs.

\subsection{Data Collection}

We collect information on kernel execution times by modifying the CLBlast autotuning utility. Our modifications collects the sample mean time (instead of the sample minimum time collected by CLBlast) and also records the compilation time. 
We follow the default sample sizes used by CLBlast: the Xgemm kernels were executed twice.
Some kernels produce incorrect mathematical results, so we screen out data for kernels that fail to match the reference kernels.

With 16 tunable parameters, the search space of the matrix-multiply kernels is too large to explore fully. Instead, CLBlast's default hand-tuned subset of kernels is explored, followed by a random selection from the kernel space. While the random selections are not standard across devices, they give good insight into the achievable performance for each device-input pair.

Time to collect the data for each device varied from just under 17 hours on the NVIDIA Quadro to over a week on the Intel HD500 and Iris Pro, both of which ran into interruptions and required re-starts. Some devices and input combinations were unable to be timed, as the tuners ran into GPU time-outs. This was especially true on the HD500 and Iris Pro. We record the number of executed variants and inputs by device in Table \ref{tab:dataset generation}. 
A postprocessing pass adds information about the performance of each kernel relative to the best and worst kernels for that device and input.

\begin{table*}
\centering
  \caption{Sample Data From Dataset}
  \label{tab:sample data from dataset}
\begin{tabular}{ccccccc}
\toprule
Device & Input & Parameters & Runtime (ms) & Slowdown vs Oracle & Speedup vs Worst \\
\midrule
Vega & (M=512, N=1024, K=256) & (0,1,32,2,8,16,128,16,8,64,0,0,1,0,8,1)  & 2.030 & 1.24 & 55.67 \\
\hline
Vega & (M=4096, N=1024, K=512) & (0,1,16,2,8,8,128,8,8,128,1,0,1,1,1,1) & 69.048 & 42.21 & 1.64 \\
\hline
Quadro & (M=1024, N=1024, K=256) & (1,1,1,1,4,4,16,4,4,16,0,0,0,0,1,1) & 0.693 & 4.41 & 107.55 \\
\hline
Quadro & (M=1024, N=1024, K=256) & (0,1,32,2,16,16,64,8,8,64,1,1,0,0,4,4) & 0.157 & 1.0 & 474.71 \\
\bottomrule
\end{tabular}
\end{table*}

In total, we collect the compilation and sample mean execution times for kernels running a variety of input calculations and executed on 5 devices spanning 4 vendors, totalling 142,782 device-input-parameter tuples. A sample of our dataset is shown in Table \ref{tab:sample data from dataset}.

\subsection{CLBlast Dataset}
\label{sec:CLBlast Dataset}
CLBlast provides a crowdsourced dataset of tuning results with data for over 100 GPU devices. 
In this crowdsourced data, (1) each kernel is only executed for a single input matrix, (2) the sample minimum time is recorded, not the sample mean, and (3) the compilation times are not stored. 
We further prune data from older versions of the CLBlast library,\footnote{For example, data entries that do not provide information about the device driver.} resulting in a dataset including performance results from 71 devices. 

Since it has data for only a single input, the CLBlast dataset does not provide information about input sensitivity. However, it does allow us to (1) check if a certain parameter configuration selected by portability tuning is valid across a larger number of devices than we have, and (2) use the data to evaluate the effect of more data on the degree of performance portability that can be achieved.

\subsection{Evaluation Methodology}

We compare portability tuning against two state-of-the-art techniques: CLBlast and Oracle. 

To compare against CLBlast's auto-tuner, we compare against the parameters that would be selected by the CLBlast library. In case the selections were not verified to produce correct results on other input sizes, we iterate until we find a selection that does.

Portability tuning is also compared against the Oracle. The Oracle represents the performance that would be achieved if the best-known configuration could be used for every execution (i.e. device + input pair). Such an exhaustive method is not practical but provides an upper-bound on performance.

We evaluate three implementations of portability tuning. The first is a metaheuristic search approach implemented with the OpenTuner autotuner~\cite{OpenTuner}, which we refer to as \textit{PortabilityTune}. The other two implementations use the decision tree and $k$-means clustering from scikit-learn \cite{scikit-learn}. Unless otherwise stated, all three methods are run 30 times, with the mean value shown. The error bars in all figures indicate 95\%  confidence intervals. The methods are run offline until convergence or 30 seconds passes, whichever occurs first.

\textbf{Figures of Merit}. We use the following metrics to evaluate the performance of different tuning implementations.

\begin{itemize}
    \item \textit{Slowdown Over Oracle, (S/O)} (lower is better). If the Oracle execution time is $O_{d,i}$ for some device $d$ and input $i$, and the execution time is $A_{d,i}$ with a different configuration, the slowdown of that configuration is $A_{d,i}$ / $O_{d,i}$. Optimal performance corresponds to slowdown of 1.0x.
    \item \textit{Fleet Rate} (higher is better). The fleet rate, as defined in Section \ref{Portability Tuning Framework}, is the rate at which a fleet of devices can complete some task. For the fleet task, tuning is biased towards the highest-performing devices, and the most-common devices and inputs.

\end{itemize}

\begin{figure}
    \centering
    \includegraphics[width=0.49\textwidth]{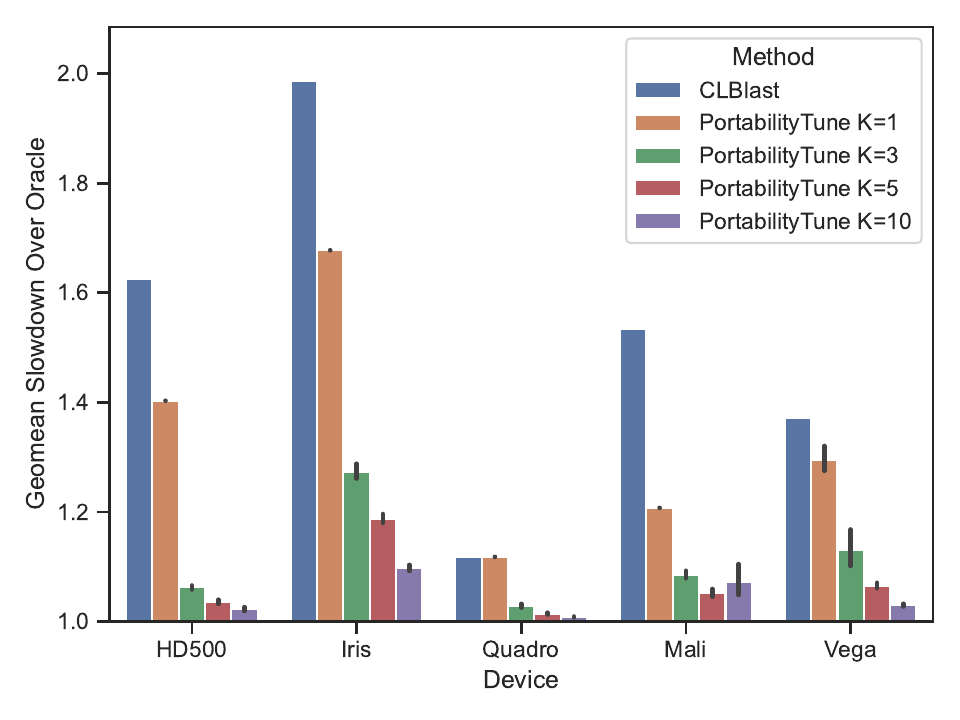}
    \caption{Device-Specific XGEMM Tuning, 5 Devices, Offline Tuning Time = 30s}
    \Description{Portability-tuned kernels outperform the default tuned CLBlast kernels. As the number of kernels to select from increases, the geomean performance increases.} 
    \label{fig:Device Specific GEMM Tunings on 5 Devices}
\end{figure}

\begin{figure}
    \centering
    \includegraphics[width=0.49\textwidth]{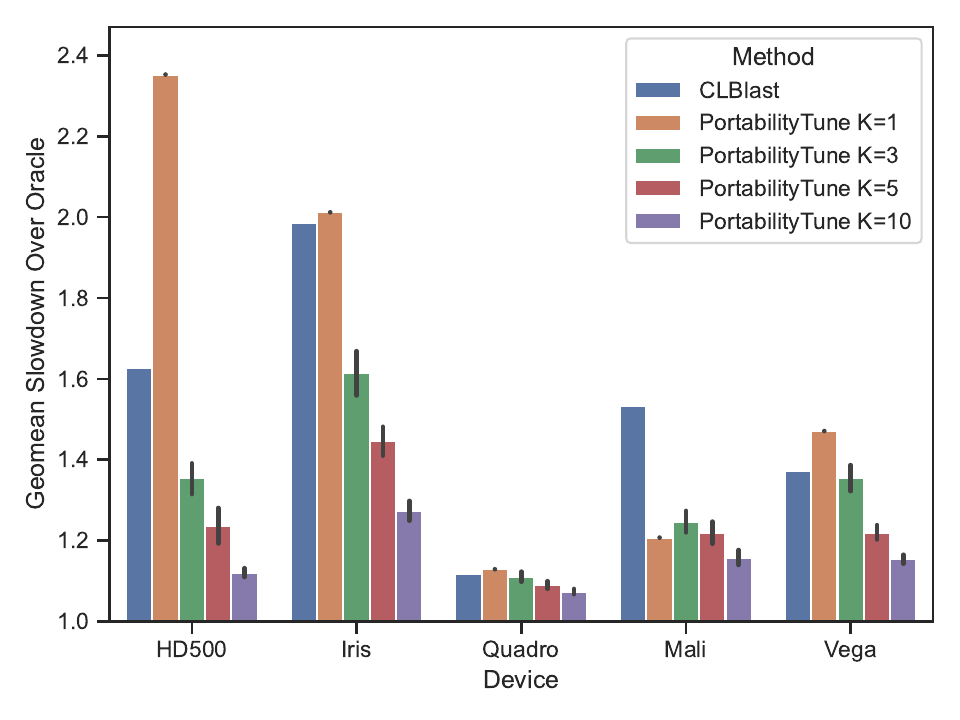}
    \caption{Device-Portable XGEMM Tuning, 5 Devices, Offline Tuning Time = 30s}
    \Description{A graph showing that as the number of kernels to select from increases, the geomean performance increases.}
    \label{fig:Device Portable Xgemm}
\end{figure}

\begin{figure}
    \centering
    \includegraphics[width=0.45\textwidth]{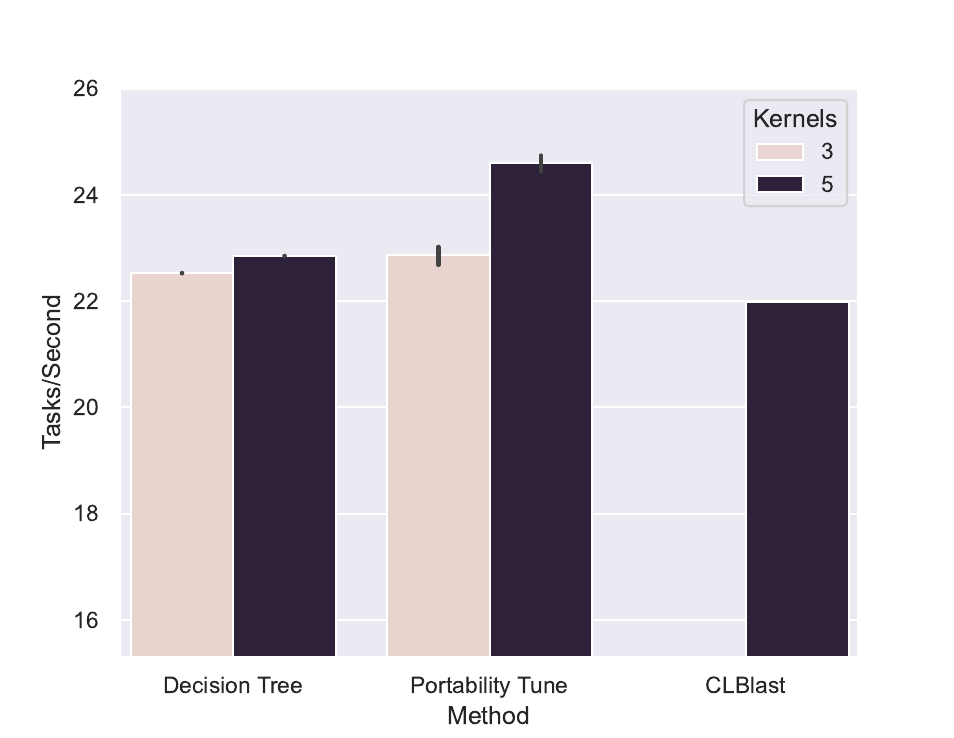}
    \caption{Median Task Rate, 5 Devices, XGEMM, Offline Tuning Time = 30s}
    \Description{A graph showing the task rate achieved by fleet tuning using different methods.}
    \label{fig:Fleet Tuning}
\end{figure}

\subsection{Comparing Portability Tuning to CLBlast} \label{Comparing Portability Tuning to the State of the Art}

We compare portability tuning to CLBlast’s default autotuning.
We examine three scenarios: 1) tuning for a specific device, 2) tuning for multiple devices, 3) tuning for fleet performance.
For these experiments, data labeled CLBlast  shows the performance of tuning using CLBlast’s default autotuning, \textit{which is always a single variant specific to the underlying device}. 
Portability tuning data is from tuning that utilizes stochastic search.

\subsubsection{Tuning Variants for a Single Device}
First, we observe that portability tuning outperforms CLBlast's autotuning when variants are tuned specific to each device. In Figure \ref{fig:Device Specific GEMM Tunings on 5 Devices}, we run PortabilityTune individually on each device, where the search space is that device and the inputs available on that device (see Table \ref{tab:dataset generation}). As shown in the figure, portability tuning just a single kernel variant provides a statistically significant performance benefit for 4 of 5 devices.

With additional variants, portability tuning becomes competitive with the Oracle (which utilizes the best-known kernel for every invocation), easily outperforming CLBlast’s autotuning. With 10 Xgemm kernel variants, the portability tuned variants perform within 1.1x the Oracle on all 5 of our devices.

\subsubsection{Tuning Variants Shared Across Devices}
Second, we observe that portability tuning is competitive to, or outperforms, CLBlast when tuning for a small set of possible devices. In this scenario, CLBlast tuned kernels are specific to each device (as in the previous figure), but portability tuning is done across the set of devices, with the chosen variants are pooled across devices. As shown in Figure~\ref{fig:Device Portable Xgemm}, portability tuning just a \emph{single} shared Xgemm variant shows competitive performance to CLBlast's autotuning on 4 of the 5 devices. 

However, the portability tuned Xgemm variant struggles on the HD500 device. On further examination, we observed that, due to the gaps in our dataset, the HD500 device had fewer collected inputs than the other devices.\footnote{45 matrix-multiplications on the HD500 as opposed to 64 on the Vega; see Table \ref{tab:dataset generation}} Because of this, only 15\% of the performance results in the fitness evaluation are from the HD500, as compared to 22\% of the Vega. This leads the HD500 to be weighed less in the fitness evaluation than the other devices.

With additional shared variants, portability tuning largely outperforms CLBlast and begins to become competitive with the Oracle. With 10 Xgemm variants, the portability tuned variants perform within 1.2x the Oracle on 4 of our 5 devices. Similarly, with 3 Xger kernel variants, portability tuning variants achieve a performance of 1.2x the Oracle for 4 of the 5 devices.

\subsubsection{Tuning Variants For Fleet Optimization}
Lastly, we observe that portability tuning outperforms CLBlast when tuning a fleet of devices.
Figure \ref{fig:Fleet Tuning} shows the median task rate for different strategies of variant selection. For this experiment, we consider running ``tasks'' on our 5 devices, where each task is completed when a matrix multiplication for available known input is completed once. For example, a task on the HD500 includes running 45 unique matrix-multiplications, and a task on the Vega includes running 64 unique matrix-multiplications. The chosen variants are shared across devices. In this case, the CLBlast technique represents sharing all 5 of the \textit{device-specific} tuned kernels across the devices.
As shown in Figure \ref{fig:Fleet Tuning}, both Decision Tree and the PortabilityTune implementation of portability tuning outperform the variants chosen by CLBlast. The PortabilityTune tuning in particular outperforms CLBlast, with an 11\% improvement in task rate (an additional 9,000 tasks per hour) for the same number of variants (5).

\subsection{Comparing Portability Tuning Implementations} \label{Comparing Portability Tuning Implementations}

\begin{figure*}
    \centering
    \includegraphics[width=1\textwidth]{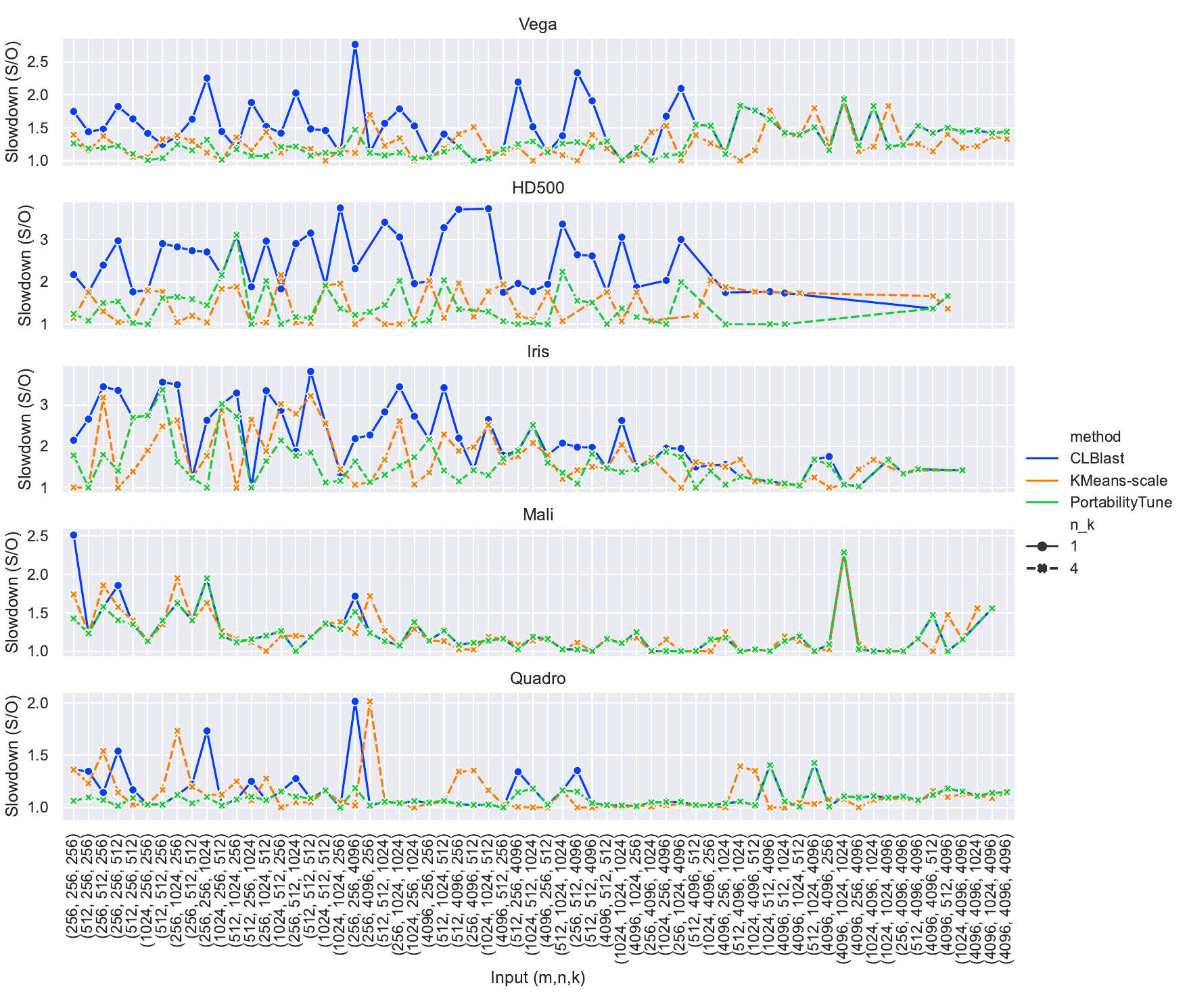}
    \caption{Implementation Comparison}
    \Description{First, offline, kernels are selected from a large space of possible kernels. These kernels cover different parts of the overall space of devices and inputs. Then, online, a classifier determines which of the selected kernels should be executed from the given device and input.}
    \label{fig:IPScatter Implementation Comparison}
\end{figure*}

In this section, we compare the different implementations of portability tuning, i.e. clustering, decision tree, and PortabilityTune. Our results show that implementations have tradeoffs: (1) the time to converge on a result, (2) performance on specific input sizes, (3) their performance on non-standard tasks, such as the fleet tuning task.

\subsubsection{Tradeoffs in Tuning Time and Convergence Performance}
Figure \ref{fig: Tuning Time Convergence, Iris Pro} compares the performance of the K-Means Clustering, Decision Tree, and PortabilityTune techniques over time for the Xgemm routine for a single tuning run. Both K-Means and Decision Tree converge quickly and terminate prior to one second of execution. This is also true for PortabilityTune with only a single variant. With more variants, the search space for PortabilityTune quickly becomes intractable.

We observe that the stochastic search implementation may take much longer to converge than the Clustering and Decision Tree implementations of portability tuning. In Figure \ref{fig: Tuning Time Convergence, Iris Pro}, we see that PortabilityTune may take several hundred seconds to converge, whereas the Decision Tree and Clustering implementations take under a second to converge. However, the maximum performance achieved by PortabilityTune for 4 and 10 variants is greater than that achieved by the other methods. This is evidence that shows that while the Clustering and Decision Tree methods are time-efficient, they may not select variants with the highest attainable performances.

\subsubsection{Per-Input Variations}
Figure \ref{fig:IPScatter Implementation Comparison} compares the median Slowdown Over Oracle of the CLBlast, K-Means Clustering, and PortabilityTune (stochastic) techniques for each device and input size in the dataset on the Xgemm routine. The selected variants are shared across the devices.

We observe that the performance may vary on a per-input basis between portability tuning techniques. In Figure \ref{fig:IPScatter Implementation Comparison}, we note that the overall performance between the K-Means Clustering and PortabilityTune techniques is similar. However, the per-input performances vary widely, in some cases by nearly 3x. 
PortabilityTune never underperforms the variants chosen by the CLBlast technique, whereas the K-Means technique occasionally underperforms the CLBlast-chosen variants.

\subsubsection{Tradeoffs in Non-Standard Tasks}
Lastly, we observe that the stochastic search implementation of PortabilityTune outperforms other implementations of portability tuning for the fleet tuning task. As shown in Figure \ref{fig:Fleet Tuning}, PortabilityTune provides statistically significant performance improvements over the Decision Tree implementation of portability tuning. On further investigation, we note that different device-input combinations have different impacts on the task rate. Overall, the task rate is biased towards common, high-rate devices. For example, improving the variants’ performance of the faster devices (eg. Vega, Quadro) leads to much greater impacts on the task rate than improving the performance of the variants on the slower devices. The Decision Tree and Clustering implementations do not directly account for this impact in their selection of variants, and thus cannot reach the highest performance gains found by the stochastic search implementation PortabilityTune.

\subsection{Generalizing To Unseen Environments}

\begin{figure}
    \centering
    \includegraphics[width=0.49\textwidth]{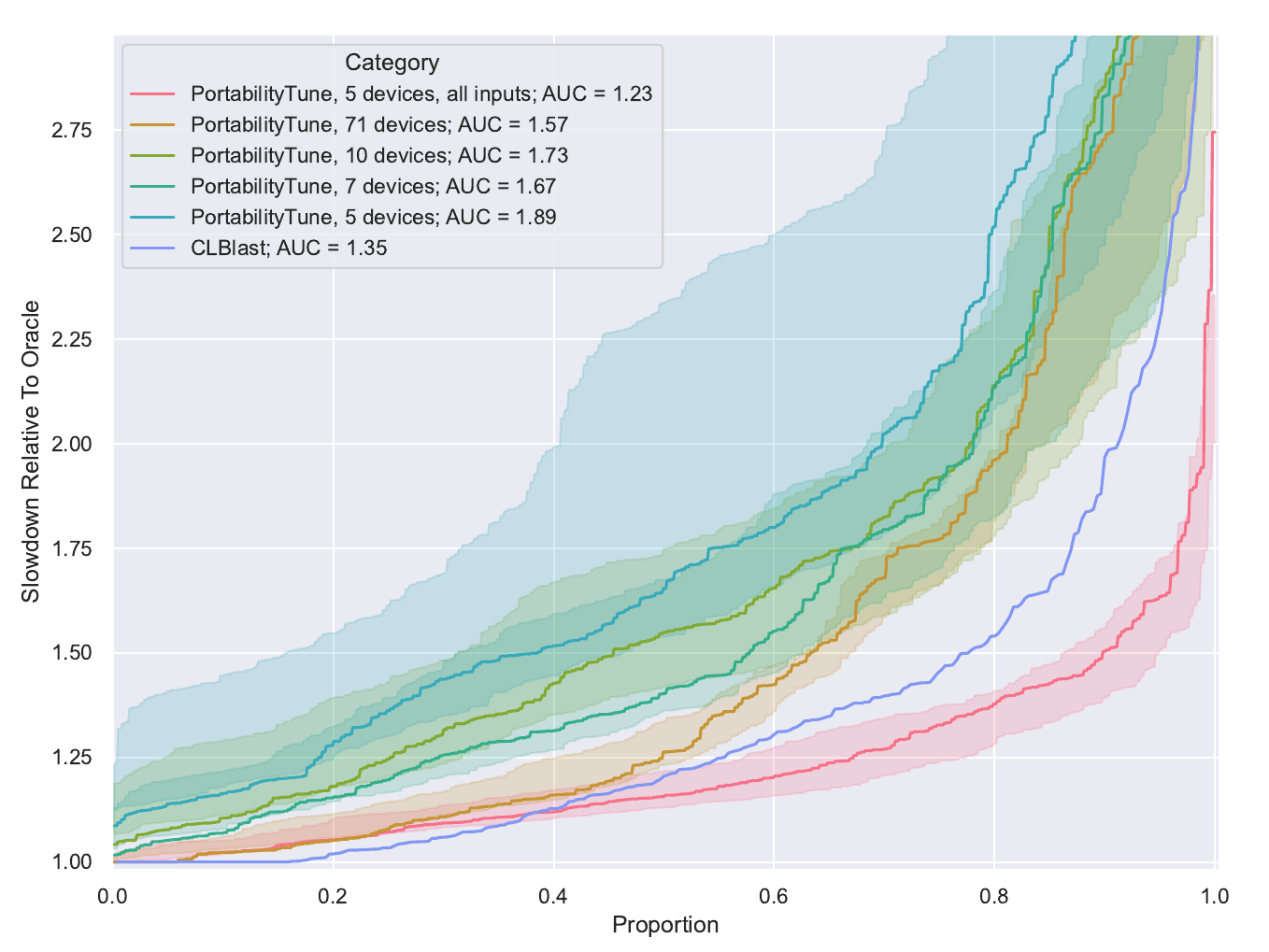}
    \caption{Portability Tuning On Unseen Devices, 5 Variants, Offline Tuning Time = 30s}
    \Description{A graph showing that much of the potential performance from a portability tuning can be achieved by exploring a fraction of the kernel parameter space.}
    \label{fig: Portability Tuning Unseen Devices K=5}
\end{figure}

\begin{figure}
    \centering
    \includegraphics[width=0.49\textwidth]{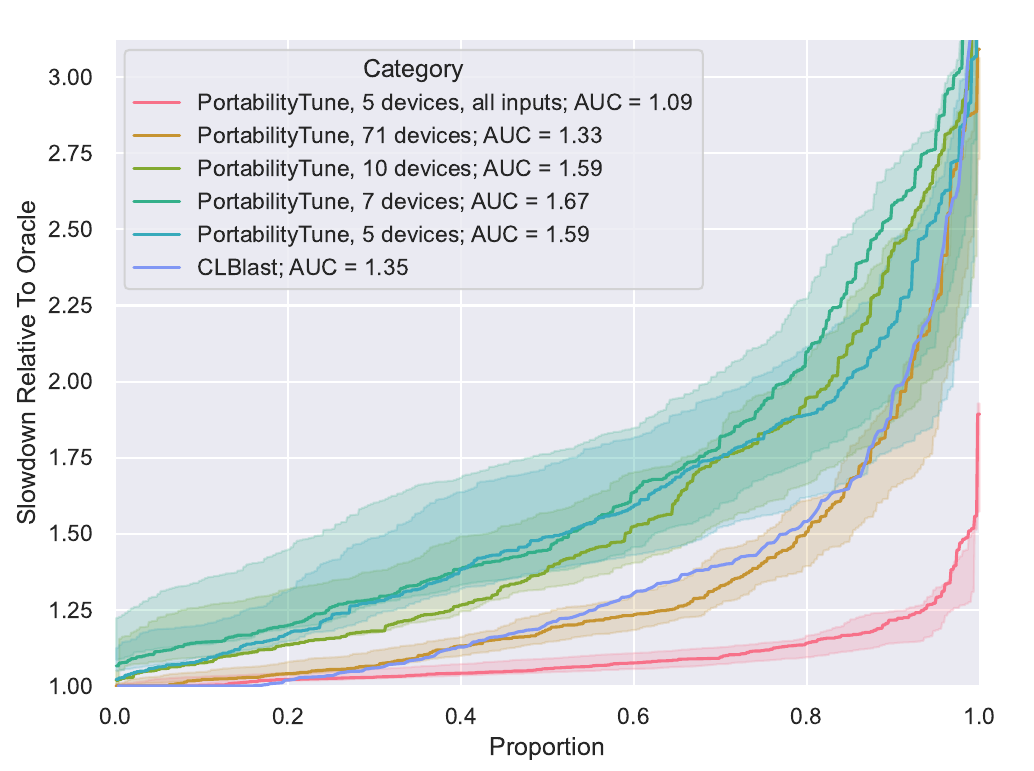}
    \caption{Portability Tuning On Unseen Devices, 15 Variants, Offline Tuning Time = 30s}
    \Description{A graph showing that much of the potential performance from a portability tuning can be achieved by exploring a fraction of the kernel parameter space.}
    \label{fig: Portability Tuning Unseen Devices K=15}
\end{figure}

Performance portable code should  be resilient to changes in the underlying device~\cite{Balaprakash, PENNYCOOKNAVIGATINGPERFPORT, Deakin2019}. 
We evaluate this characteristic by examining how well portability tuned code generalizes onto unseen devices. 
Our results show that portability tuned code can generalize to unseen devices and is competitive to re-tuning.

To evaluate how portability tuned variants would perform on unseen devices, we PortabilityTune using the data from varying numbers of randomly selected devices $|D|$ from the CLBlast dataset described in Section \ref{sec:CLBlast Dataset} (from which there are 71 devices). Then, we evaluate the performance of the selected variants on the disjoint 292 device-input combinations in our dataset. We compare against two baselines: (1) we PortabilityTune for the 292 device-input combinations in our dataset, and (2) the CLBlast-selected variants for the 5 devices. It is worth noting the CLBlast dataset does not contain any input variation, using only a single input size. The selected variants are shared among the devices.

We show the result of this experiment in Figures \ref{fig: Portability Tuning Unseen Devices K=5} and \ref{fig: Portability Tuning Unseen Devices K=15}, selecting 5 and 15 variants, respectively.\footnote{The CLBlast method selects 5 variants in both figures, one tuned variant per device.} The figures show the proportion of device-input combinations achieving some Slowdown over Oracle for each technique of selection. Thirty runs are performed for the stochastic search technique PortabilityTune. Of these runs, we show the median-performing run and a 95\% confidence interval for the performance the variants on each device-input combination. 

\subsubsection{The Impact of Devices on Performance Generalization}
We observe several findings from the experiment. 
First, the number of devices for which PortabilityTune is tuned on significantly affects how well the selected variants generalize. With 5 (of 71) devices known to the tuning process, a slowdown of 1.25x or less is achieved on roughly 20\% of the device-input combinations (for both 5 and 15 variants selected). In contrast, when tuning using all 71 devices from the CLBlast database, a slowdown of 1.25x or less is achieved on 50\% and 64\% of the device-input combinations, respectively.

\subsubsection{Achieving Autotuning Performance Without Autotuning}
Second, we find that PortabilityTune achieves comparable performance to the CLBlast-selected variants. For $|K|=5$ the performance between CLBlast and PortabilityTune is comparable -- though a performance gap does exist, particularly with respect to the proportion of low-performing devices. For $|K|=15$, the variants selected by PortabilityTune with 71 devices slightly outperforms those found by CLBlast's autotuning.
It is worth noting that PortabilityTune on unseen devices and CLBlast similarly lack knowledge of the variety of inputs in the evaluation space. That the performance between the techniques is comparable provides evidence that the code produced by portability tuning is resistant to changes in the underlying device. This, in turn, is strong evidence of the performance portability of variants selected by portability tuning. It also shows that the performance of traditional autotuning is achievable \textit{without} autotuning for a new device.

\subsubsection{The Value of Tuning on Known Devices}
Lastly, we observe that a significant performance gap exists between PortabilityTune with unseen devices as compared to PortabilityTune with the devices known (1.57x vs. 1.23x for $|K|=5$ and 1.33x vs. 1.09x for $|K|=15$.) This provides evidence that portability tuning for known devices/inputs remains valuable and can provide additional performance over variants originally portability tuned for other devices.

\subsection{Limitations of Portability Tuning}
\label{Limitations}

\begin{figure}
    \centering
    \includegraphics[width=0.49\textwidth]{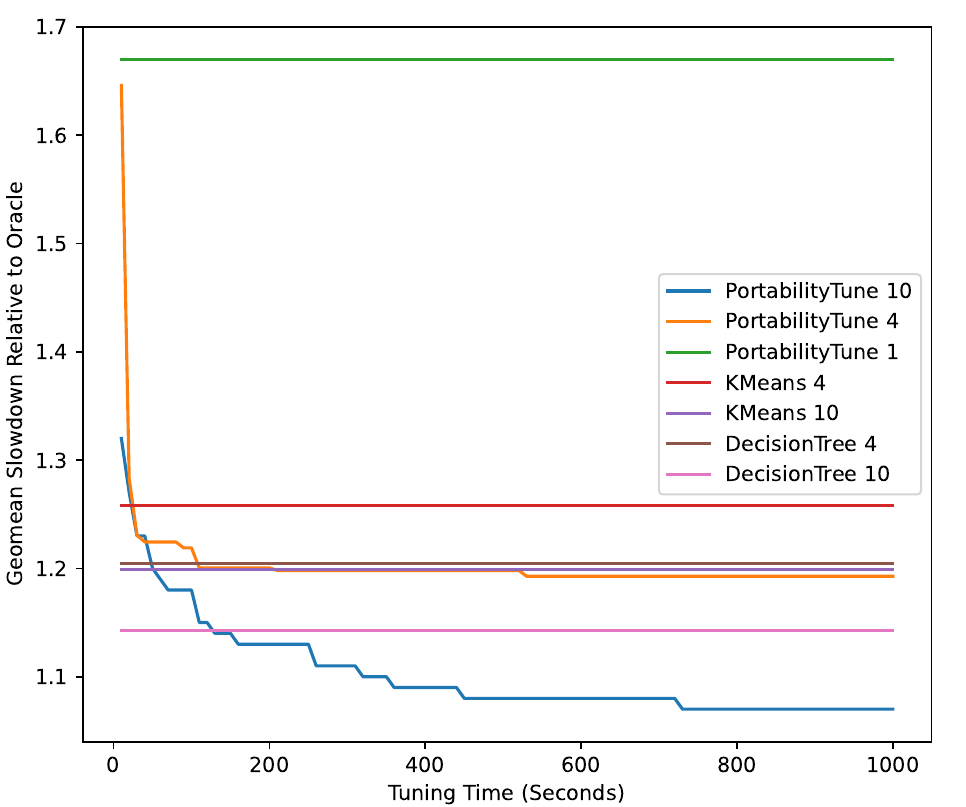}
    \caption{GEMM Performance vs Tuning Time, Iris Pro, All Inputs}
    \Description{A graph showing that increased number of selected kernels increases the time to select near-optimal kernels for a set of that size.}
    \label{fig: Tuning Time Convergence, Iris Pro}
\end{figure}

The cost of Portability tuning is its primary limitation  as discussed in Section \ref{Comparing Portability Tuning Implementations}. 
Additionally, portability tuning also does not provide a clear way to generalize the selected parameters to \textit{unseen} inputs. 
The sets of kernels selected by portability tuning do not follow a known pattern that can be used to determine which kernels should be used for unseen devices, inputs, and environments. A significant body of work has created classification systems that could be used for this inference, such as works by Lawson and Goli \cite{LAWSON2021102813} and Muralidharan et al. \cite{Muralidharan}. Future work may be necessary to confirm that these systems would be effective in generalizing portability-tuned variants to unseen devices, inputs, and environments.

\section{Conclusion}
\label{sec:conclusion}

Portability tuning is a novel framework that uses experimental data to generate multi-versioned kernels with high performance portability. We evaluate portability tuning against the state-of-the-art and show that it significantly closes the gap to the theoretical maximum performance. Portability tuned programs are even portable to new execution environments with performance comparable to autotuning on those devices.

\section{Acknowledgments}
\label{sec:acknowledgments}

Robert Hochgraf was supported by multiple NSF REUs. This material is based upon work supported by the National Science Foundation under Grant \#\href{https://www.nsf.gov/awardsearch/showAward?AWD_ID=2144384}{2144384}. Any opinions, findings, and conclusions or recommendations expressed in this material are those of the author(s) and do not necessarily reflect the
views of the National Science Foundation.

\section{Disclosures}
\label{sec:disclosures}

The generative AI tools Github Copilot and GPT-4o were used to assist with coding \LaTeX~graphs, figures, and data collection scripts.


\bibliographystyle{IEEEtranS}
\bibliography{refs}

\end{document}